\documentclass[
 manuscript,
 ]{aastex631}
\usepackage{amsmath}

\submitjournal{ApJ}

\shorttitle{Electron Firehose Instabilities}
\shortauthors{Moya et al.}

\begin{document}

\title{Comparing the counter-beaming and temperature anisotropy driven aperiodic electron firehose instabilities in collisionless plasma environments}

\author[0000-0002-9161-0888]{Pablo S. Moya}
\email{pablo.moya@uchile.cl}
\affiliation{Departamento de F\'isica, Facultad de Ciencias, Universidad de Chile, Santiago, Chile}
\affiliation{Centre for mathematical Plasma Astrophysics, KU Leuven, Celestijnenlaan 200B, B-3001 Leuven, Belgium}

\author[0000-0003-3223-1498]{Rodrigo A. L\'opez}
\affiliation{Departamento de F\'{\i}sica, Universidad de Santiago de Chile, Usach, 9170124 Santiago, Chile}

\author[0000-0002-8508-5466]{Marian Lazar}
\affiliation{Centre for mathematical Plasma Astrophysics, KU Leuven, Celestijnenlaan 200B, B-3001 Leuven, Belgium}
\affiliation{Institut f\"{u}r Theoretische Physik, Lehrstuhl IV: Weltraum- und Astrophysik, Ruhr-Universit\"{a}t Bochum, D-44780 Bochum, Germany}

\author[0000-0002-1743-0651]{Stefaan Poedts}
\affiliation{Centre for mathematical Plasma Astrophysics, KU Leuven, Celestijnenlaan 200B, B-3001 Leuven, Belgium}
\affiliation{Institute of Physics, University of Maria Curie-Skłodowska, Lublin, Poland}

\author[0000-0003-0465-598X]{Shaaban M. Shaaban}
\affiliation{Theoretical Physics Research Group, Physics Department, Faculty of Science, Mansoura University, 35516, Mansoura, Egypt}



\begin{abstract}
The electron firehose instabilities are among the most studied kinetic instabilities, especially in the context of space plasmas, whose dynamics is mainly controlled by collisionless wave-particle interactions. This paper undertakes a comparative analysis of the aperiodic electron firehose instabilities excited either by the anisotropic temperature or by the electron counter-beaming populations. Two symmetric counter-beams provide an effective kinetic anisotropy similar to the temperature anisotropy of a single (non-drifting) population, with temperature along the magnetic field direction larger than that in perpendicular direction. Therefore, the counter-beaming plasma is susceptible to firehose-like instabilities (FIs), parallel and oblique branches. Here we focus on the oblique beaming FI, which is also aperiodic when the free energy is provided by symmetric counter-beams. Our results show that, for relative small drifts or beaming speeds ($U$), not exceeding the thermal speed ($\alpha$), the aperiodic FIs exist in the same interval of wave-numbers and the same range of oblique angles (with respect to the magnetic field direction), but the growth rates of counter-beaming FI (CBFI) are always higher than those of temperature anisotropy FI (TAFI). For $U/\alpha > 1$, however, another electrostatic two-stream instability (ETSI) is also predicted, which may have growth rates higher than those of CBFI, and may dominate in that case the dynamics.
\end{abstract}



\section{Introduction} \label{sec:intro}

The electron firehose instabilities are among the most studied kinetic instabilities in plasmas, especially in the context of space plasmas, whose dynamics is mainly controlled by the wave-particle interactions rather than particle-particle collisions \citep{Li-Habbal-2000, Gary-Nishimura-2003, Camporeale-Burgess-2008, Shaaban-etal-2019, Lopez-etal-2019, Saeed-etal-2017,Shaaban-etal-2018a, Shaaban-etal-2018b, Lopez2020}. These instabilities are generally referred as produced by the free energy stored in the form of electron temperature anisotropy $A_e \equiv T_{e,\perp}/T_{e, \parallel} < 1$ (with $\perp,\parallel$ denoting gyrotropic directions with respect to the magnetic field) \citep{Li-Habbal-2000, Gary-Nishimura-2003,Camporeale-Burgess-2008, Shaaban-etal-2019}. In this case, linear theory predicts two branches, a periodic electron firehose instability, i.e., with finite wave frequency $\omega_r \ne 0$, and propagating parallel and at small angles (quasi-parallel) with respect to the magnetic field, and an aperiodic branch ($\omega_r = 0$) developing only for oblique angles, and  much faster than the periodic one \citep{Li-Habbal-2000,Camporeale-Burgess-2008, Shaaban-etal-2019}. However, firehose-like instabilities are also expected to be driven by other forms of free energy, namely, electron populations drifting or beaming along the magnetic field \citep{Saeed-etal-2017, Shaaban-etal-2018a, Shaaban-etal-2018b, Lopez2020}. Thus, the periodic (quasi-parallel) branch can be also excited, as a firehose heat-flux instability triggered by the anti-sunward drift (or beaming speed) of the suprathermal electron populations (i.e., halo,  strahl), which carry the main heat flux in the solar wind plasma \citep{Saeed-etal-2017, Shaaban-etal-2018a, Shaaban-etal-2018b}.

\cite{Lopez2020} have recently shown that in a magnetized plasma system, two electron populations counterbeaming along the uniform magnetic field may also excite an aperiodic instability, very similar to the aperiodic firehose instability driven by the anisotropic temperature of electrons \citep{Li-Habbal-2000, Gary-Nishimura-2003, Camporeale-Burgess-2008, Shaaban-etal-2019, Lopez-etal-2019}. The aperiodic nature of beaming instability is highly conditioned by the symmetry of counterbeaming electrons, which are similar to an excess of temperature (equipartitioned kinetic energy) in parallel direction, i.e., $T_{e,\perp} < T_{e, \parallel}$ \citep{Lopez2020, Kim2020}. Moreover, both aperiodic instabilities develop obliquely to the magnetic field (${\bf k} \times {\bf B}_0 \ne 0$). If the symmetry of counterbeams (e.g., in particle density or temperature) is altered the instability converts to a periodic mode, still propagating obliquely to the magnetic field. Despite these similarities, the nature and amount of free energies triggering the instabilities are not necessarily the same, and may therefore determine a series of contrasting properties which are investigated in the present paper. Such an investigation is not only of theoretical but also of practical interest, such as for the analysis of electron velocity distributions observed in space plasmas, e.g., on opened or closed magnetic field topologies associated with interplanetary shocks \citep{Wimmer-2006, Skoug-etal-2006, Lazar-etal-2014}. Thus, in the electron velocity distributions one may often need to distinguish  between field-aligned counter-beams and an excess of temperature along the magnetic field \citep{Wimmer-2006, Lazar-etal-2014}, or even a depletion of pitch-angle distributions (PADs) at 90$^{\rm o}$ that can arise along open magnetic field lines \citep{Gosling-etal-2001} \footnote{Suprathermal electron 90$^{\rm o}$ PAD depletions can result from focusing and mirroring of electrons in connection to magnetic field enhancements upstream of interplanetary shocks or beyond the spacecraft \citep{Gosling-etal-2001, Skoug-etal-2006}.}. In these cases PADs show similar anisotropic profiles, with field aligned peaking densities, i.e., at 0$^{\rm o}$ and 180$^{\rm o}$ (for both closed and open magnetic field topologies), and a differentiation between different kind of electron anisotropy may be supported by distinguishing between the corresponding self-induced wave fluctuations.

In the present paper we report the results of a comparative study of these two instabilities, the counter-beaming firehose instability (CBFI) and the temperature anisotropy firehose instability (TAFI). Our study is intended to an extended characterization of their similarities, but also contrasting features, which are directly conditioned by the zero-order conditions of the plasma system, namely, the initial velocity distributions of electrons. 
The analysis is focused on linear properties of the unstable modes, like, growth rates and their maximums associated with the fastest growing modes, the range of unstable wave-numbers, and angles of propagation. In the next section we present the two kinetic models, and introduce the so-called \emph{effective anisotropy}~\citep{Hadi2014}, a parameter that allows us to quantify the free energy driving the CBFI and stored in the differential streaming of the beams, and to directly compare with TAFI. Then, in section~\ref{u-spectra} we perform a comprehensive linear analysis of the instabilities, computing growth rates and their variations with propagation angle, wave number, plasma beta, and anisotropy (temperature anisotropy for the TAFI and effective anisotropy for the CBFI). A more contrasting examination of these two instabilities includes the interplay of CBFI with the electrostatic two-stream instability (ETSI), if the beaming velocity ($U$) of the counter-beams is comparable or even exceeds the thermal speed ($\alpha$); see section~\ref{etsi}. The study is completed in section~\ref{sec:thres} with a final comparative analysis of the instability thresholds derived in terms of anisotropy and electron (parallel) plasma beta for both TAFI and CBFI. In the last section we summarize our main results and present conclusions. 

\section{Models}
\label{sec:theory}
%

We consider two distinct collisionless quasi-neutral electron-proton plasma systems, to be compared in the next. One of them is assumed with non-drifting electrons but with an intrinsic temperature anisotropy 
\begin{equation}
    A\equiv T_{\perp}/T_{\parallel}<1, \label{ta}
\end{equation} described by a  bi-Maxwellian distribution
\begin{equation}
f_e(v_\perp,v_\parallel)=\frac{n_e}{\pi^{3/2}\alpha_{\perp}^2 \alpha_{\parallel}}
\exp\left(-\frac{v_\perp^2}{\alpha_\perp^2}
-\frac{v_\parallel^2}{\alpha_\parallel^2}\right)\,.
\label{eq:bimax}
\end{equation}
Here $n_e$ is the electron total number density, $\alpha_{\perp, \parallel} =~\sqrt{2k_BT_{\perp, \parallel}/m_e}$ are the thermal velocities of electrons in perpendicular ($\perp$) and parallel ($\parallel$) directions to the background magnetic field. This expression of temperature anisotropy in Eq.~\ref{eq:bimax} is widely used in the literature \citep{Li-Habbal-2000, Gary-Nishimura-2003, Lopez-etal-2019, Shaaban-etal-2018a}, and for conditions of interest here, i.e., $T_\parallel > T_\perp$, lower (sub-unitary) values of anisotropy ($A < 1$) signify larger deviations from isotropy ($A = 1$).

The second plasma system comprises two electron populations counter-moving along the magnetic filed (in a rest frame fixed to protons), and each of them assumed with isotropic temperatures, well represented by a superposition of two drifting Maxwellian functions
\begin{equation}
f_e(v_\perp,v_\parallel)=\sum_{j=1,2} \frac{n_j}{\pi^{3/2}\alpha^3_j}
\exp\left[-\frac{v^2_\perp +(v_\parallel-V_j)^2}{\alpha^2_j}\right]\,.
\label{eq:beams}
\end{equation}
Here $n_j$ is the number density, $\alpha_j=~\sqrt{2k_BT_j/m_e}$ are the thermal velocities, and $V_j$ the drift velocity of the $j$-th beam. We also assumed fully symmetric counter-beaming electrons with the same number densities $n_1=n_2=0.5\, n_e$, same drifts (or beaming speeds) $|V_1|=|V_2|=U$ and same intrinsic temperatures $T_1 = T_2$ (in a frame fixed on protons). Thus, the zero current condition is maintained by $V_2=-n_1/n_2\,V_1 = -V_1$. Also, in both plasma systems protons are assumed isotropic and described by Maxwellian velocity distribution, i.e.,  Equation~(\ref{eq:bimax}) with $\alpha_{\parallel}=\alpha_{\perp}$ and total density $n_p = n_e$, so that the plasma is quasi-neutral. 

If the electron counter-beams are symmetric enough, one can define an effective temperature anisotropy~\citep{Hadi2014} for the second plasma system, as the ratio between the average kinetic energy of all electrons perpendicular to the magnetic field, and the average kinetic energy along the field (both computed in the proton frame). Applying this definition to our models above, the effective anisotropy reads as 
\begin{equation}
A_{\rm{eff}}=\sum_{j=1,2} \frac{n_j}{n_e} \frac{\alpha^2_j}{\alpha^2_j+2 V^2_j} = \left(1+\frac{2\, U^2}{\alpha^2}\right)^{-1} < 1.
\label{aeff}
\end{equation}
From this expression we observe that $A_{\rm{eff}} < 1$, as long $U \ne 0$, and that the level of effective anisotropy is determined by the ratio between the relative drift and the thermal speed of the beams ($U/\alpha$), such that $A_{\rm{eff}}$ decreases for increasing $U/\alpha >0$. The isotropy ($A_{\rm{eff}} = 1$) is obtained only if $U=0$. It is also worth to mention that if $U/\alpha = 1$ then $A_{\rm{eff}} = 1/3$. Thus, any particular value of $A_{\rm{eff}} < 1/3$ will correspond to $U/\alpha > 1$, and it is expected the plasma to become unstable to the electrostatic two-stream instability  (ETSI)~\citep{Lopez2020}.
%
\begin{figure}[ht!] 
  \centering
    \includegraphics[width=0.9\textwidth]{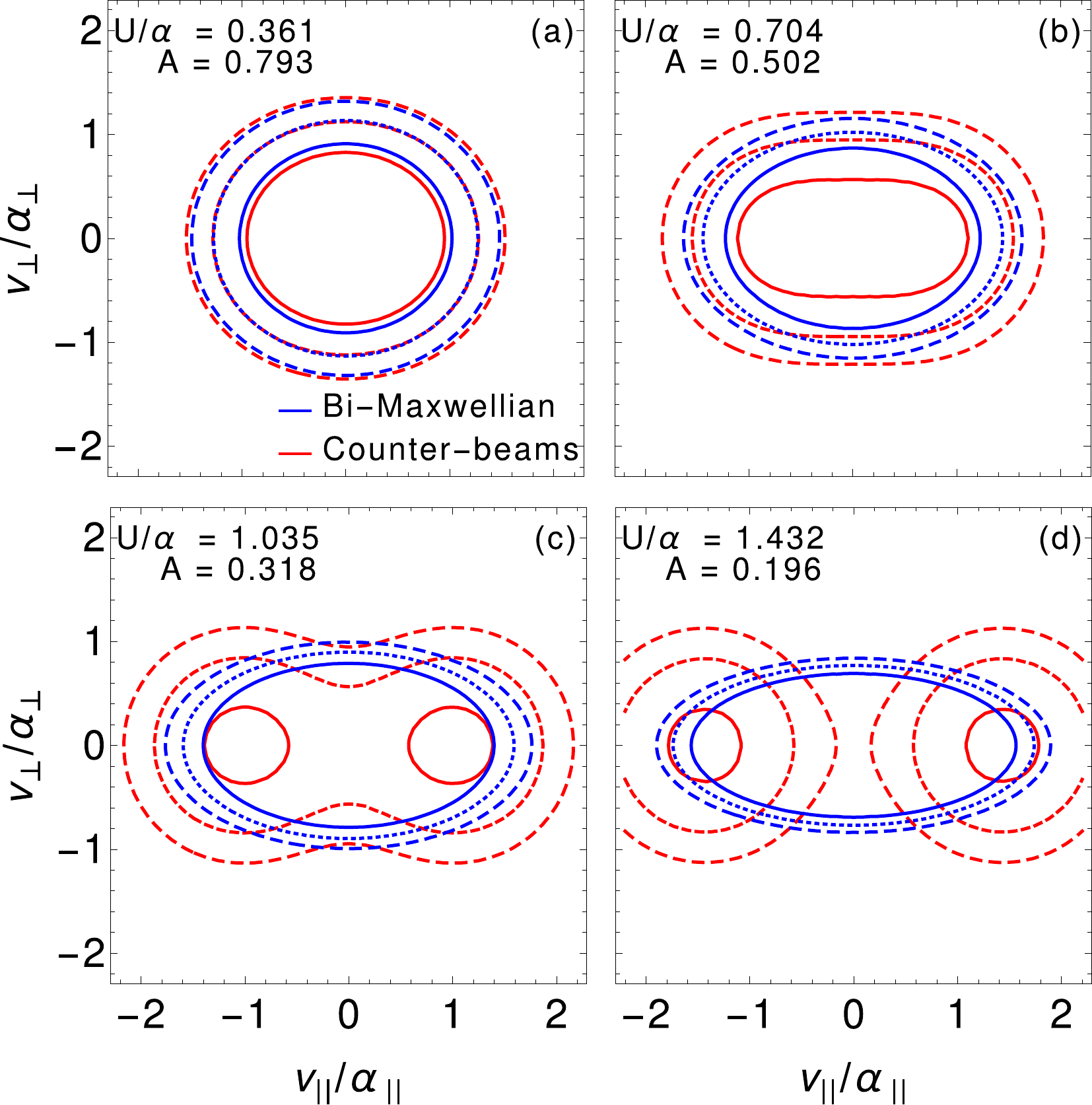}
    \caption{\label{vdf} VDF contours for the bi-Maxwellian (blue) and counter-beaming (red) electrons. Solid, dotted, and dashed lines correspond to the contour levels $\log(f_e) = -1.1$, $\log(f_e) = -1.3$, and $\log(f_e) = -1.6$, respectively. In each panel the drift $U$ of the beams is selected so that the effective anisotropy is equal to the temperature anisotropy of the bi-Maxwellian case. Namely: (a) $U/\alpha =0.36$ or $A_{\rm{eff}} = 0.79$, (b) $U/\alpha =0.70$ or $A_{\rm{eff}} = 0.50$, (c) $U/\alpha =1.04$ or $A_{\rm{eff}} = 0.32$, and (d) $U/\alpha =1.43$ or $A_{\rm{eff}} = 0.20  $.
    }
\end{figure}
The use of the effective anisotropy parameter allows us to directly compare the effectiveness of CBFI with that of TAFI when $A_{\rm{eff}} \simeq A$. Figure~\ref{vdf} shows by comparison contours levels of the electron velocity distributions for the two plasma systems, the bi-Maxwellian (blue) and the counter-beaming drifting-Maxwellian (red) distribution functions, with the same anisotropies $A = A_{\rm eff}$. In all four panels the value of relative drift divided by the thermal speed $U/\alpha$ is such that both anisotropies in Equations~(\ref{ta}) and (\ref{aeff}) are equal ($A_{\rm{eff}} = A$). In each of the four panels, for both the bi-Maxwellian or counter-beaming electrons,  contours represent different velocity distribution levels, $\log(f_e) = -1.1$ (solid), $\log(f_e) = -1.3$ (dotted), and $\log(f_e) = -1.6$ (dashed). 

From Equation~(\ref{aeff}) we note that an effective anisotropy $A_{\rm{eff}}$ = 0.79 corresponds to $U/\alpha = 0.36$. In such a case, Figure~\ref{vdf}(a) indicates only a minimal, insignificant difference between contours of bi-Maxwellian  and counter-beaming electrons, and we expect both distributions to be unstable to the firehose instability, if plasma $\beta$ is large enough. 
However, this similarity is not  maintained if $U/\alpha$ increases and deviation from temperature isotropy increases (i.e., $A$ decreases) to the same extent. For instance, for $U/\alpha \sim 0.70$ (or $A_{\rm{eff}} \sim 0.50$)  Figure~\ref{vdf}(b) shows that the two distributions begin to differentiate from each other, with the one with counter-beams being more elongated in the parallel direction than bi-Maxwellian. 
Then, at the $U/\alpha \sim 1.0$ limit ($A_{\rm{eff}} = 0.32$) in panel (c) it is already possible to recognize two distinct peaks, corresponding to the counter-beams. In this case the relative drift becomes comparable with the mean thermal spread (or thermal velocity). Therefore, besides the effective anisotropy, the separation of the beams can also provide free energy to trigger the electrostatic two-streams instability by Landau non-resonant interaction~\citep{gary1993}.
It is thus expected that both, the CBFI and the ETSI to coexist. Finally, Figure~\ref{vdf}(d) shows that in order to have an effective anisotropy $A_{\rm{eff}} \sim 0.20$, the drift speed must be $U \approx 1.40 \,\alpha$. Consequently, the counter-beams are markedly separated in velocity space and the ETSI is expected to dominate. 

From this graphical analysis of velocity distributions we should expect that, depending on the levels of temperature anisotropy and the effective anisotropy, the two instabilities, TAFI and CBFI, will share many similarities. Both being firehose-type instabilities, their properties should also depend on the level of plasma magnetization, or the strength of the background (uniform) magnetic field $B_0$, whose influence (relative to the kinetic energy of the plasma particles) is embedded in the plasma beta parameter, $\beta_j=8\pi n_ek_BT_j/B_0^2$, (where $k_B$ is the Boltzmann constant). In particular, the TAFI is expected to be unstable for $A<1$ and $\beta>1$ \citep{Li-Habbal-2000, Camporeale-Burgess-2008}, and for this instability the anisotropy thresholds can also be derived, as mainly depending on the electron beta parameter \citep{Shaaban-etal-2019}. 
However, as already mentioned, in a plasma with counter-beaming electrons the CBFI can release electromagnetic (and also electrostatic) energy from the effective anisotropy, including the gradients of the distribution between the beams.
Therefore, the unstable wave spectrum of CBFI, as a function of propagation angle and wave number, as well as the thresholds in the parameter space (e.g., $A_{\rm{eff}}$ vs. $\beta$), are expected to be similar but not the same with the corresponding properties of TAFI.

\section{Comparison of the unstable spectra}
\label{u-spectra}



%
\begin{figure*}[t!]
  \centering
    \includegraphics[width=0.95\textwidth]{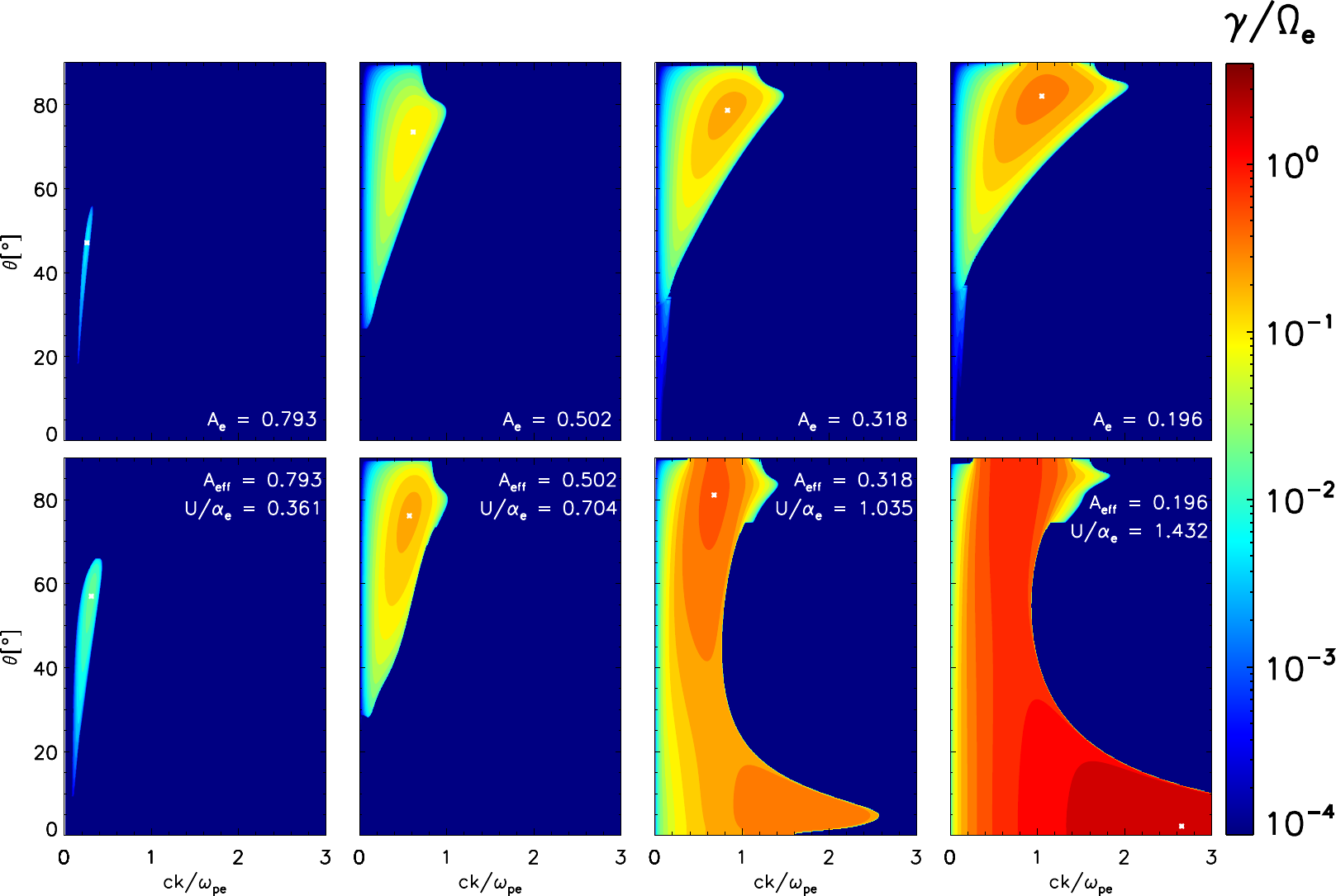}
    \caption{\label{fixed-beta} Growth rate as function of wavenumber and propagation angle for fixed beta ($\beta_e = 7.06$) and different  anisotropies, corresponding to the distributions in Figure~\ref{vdf}. Top and bottom panels correspond to TAFI and CBFI, respectively. }
\end{figure*}
\begin{figure*}[t!] 
\centering
\includegraphics[width=0.95\textwidth]{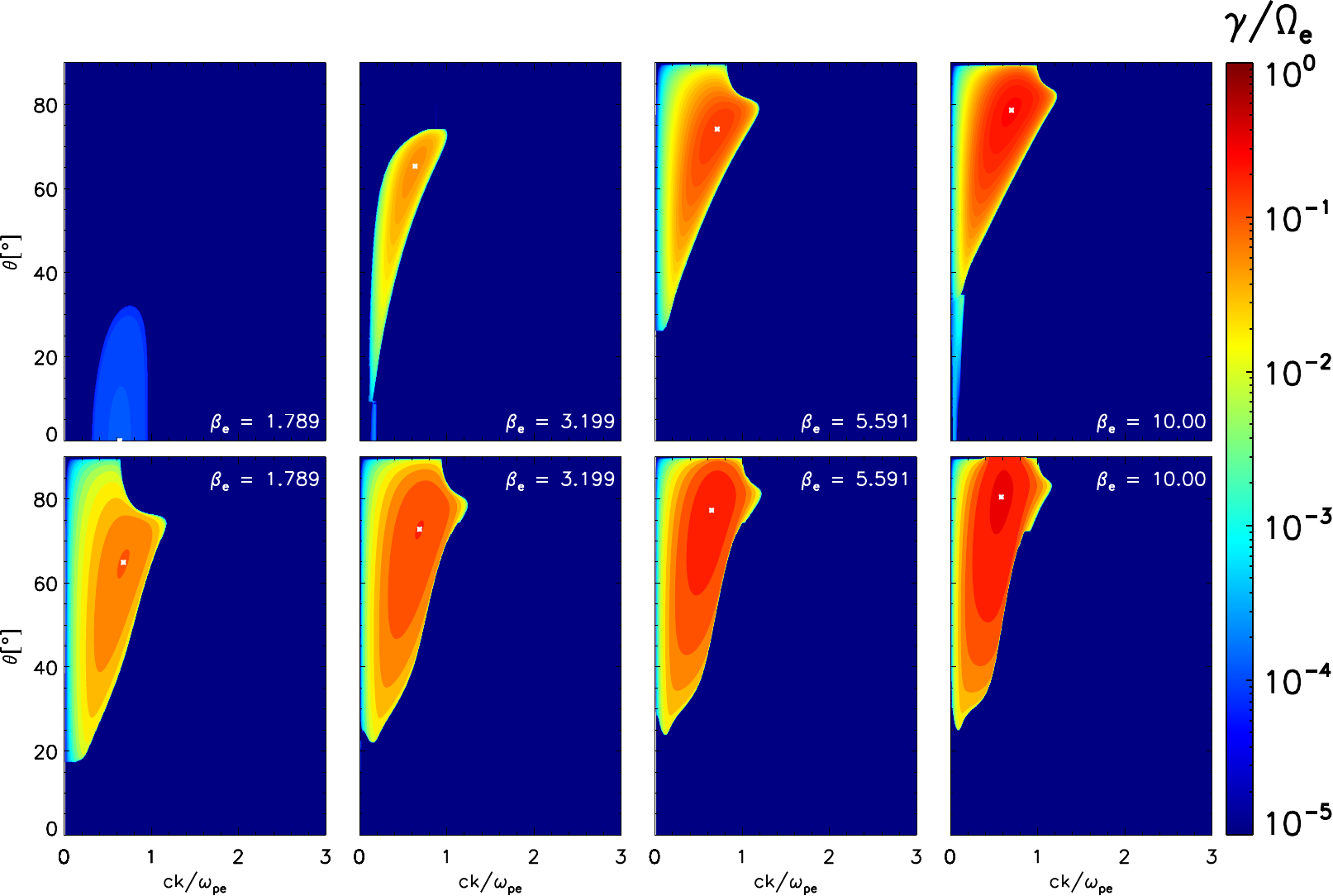}
    \caption{\label{fixed-aniso}Growth rate as function of wavenumber and propagation angle for fixed anisotropy ($A_{\rm{e}} = A_{\rm{\rm{eff}}} = 0.41$, corresponding to $U/\alpha_e = 0.85$), and different beta values. Top and bottom panels correspond to TAFI and CBFI, respectively}
\end{figure*}

In this section we present the first results of our study aiming to show how the similarities and differences in the VDFs are reflected by the dispersion and stability properties of the excited waves. 
Dipersion relations specific to each system are obtained by substituting the corresponding dispersions functions, from, e.g., Eqs.~\ref{ta} and \ref{eq:beams}, into the general dispersion equations given explicitly in textbooks \citep{gary1993, Stix1992}. 
We use the general linear dispersion solver DIS-k in the Maxwellian limit~\citep{Lopez2021} to compute the growth rate as a function of the propagation angle $\theta$ and the normalized wavenumber $ck/\omega_{pe}$ for both instabilities, see, e.g., Figures~\ref{fixed-beta} and \ref{fixed-aniso}. Table~\ref{t1} shows the values or ranges of values of the plasma parameters used in our analysis. In addition, it is important to mention that when referring to a given value of the electron beta, $\beta_e$, we mean the parallel beta ($\beta_e = \beta_{\parallel e}$) in the case of the bi-Maxwellian distribution driving the TAFI, or the total beta of both beams ($\beta_e = \beta_1+\beta_2$) in the case of the counter-beaming distribution triggering the CBFI.

\begin{table}[t!]
    \caption{Plasma parameters used in the present study$^*$\label{t1}}
    \footnotesize
    \begin{tabular}{@{}llll}
    \hline
    species (j) & bi-Maxwellian electrons & counter-beaming electrons & protons\\
    \hline
        $n_j/n_e$  & 1.0 & 0.5 & 1.0\\
        $m_p/m_j$ & 1836 & 1836 & 1.0 \\
		$T_{\perp j}/T_{\parallel j}$ & [0.1, 1.0] & 1.0 & 1.0\\
		$U_j/\alpha_j$ & 0.0 & [0.0, 2.121] & 0.0\\
    \hline 
    \end{tabular}\\
    {\footnotesize $^*$In all calculations $\omega_{pe}/\Omega_e=200$, and $1\leqslant \beta_j \leqslant 10$ for all species.}
\end{table}
\normalsize


Figure~\ref{fixed-beta} displays by comparison the growth rates (color coded) derived for a fixed value of the electron beta, $\beta_e=7.06$, and various anisotropies, for both instabilities, TAFI (upper panels) and CBFI (lower panels). For both cases the anisotropy increases from left to right. The first panels (left) show the case for $A=A_{\rm{eff}}=0.79$, which for the CBFI correspond to a drift $U/\alpha_e=0.36$. In this case, the velocity distributions for the bi-Maxwellian and counter-beaming electrons do not differ much, being similar to the case in Figure~\ref{vdf}~(a), and then only some minor differences are expected. In the upper panel the growth rate for the TAFI reaches its maximum $\gamma_{\rm{max}}/\Omega_e \approx 0.01$ at $\theta \approx47^\circ$ and $ck/\omega_{pe} \approx 0.26$, while in the bottom left panel the CBFI covers a wider range of angle and wavenumber, with a maximum of $\gamma_{\rm{max}}/\Omega_e\approx 0.02$ at $\theta\approx57^\circ$ and $ck/\omega_{pe}\approx0.31$. In the middle left panels we show the case for $A=A_{\rm{eff}}=0.50$, for which both instabilities reach larger growth rates and shift to larger angles of propagation. Again, the CBFI has a larger maximum growth rate $\gamma_{\rm{max}}/\Omega_e\approx0.32$, at a higher angle $\theta\approx76^\circ$. This results are consistent with Figure~\ref{vdf}, were it is shown that the counter-beaming electron distribution is more elongated than the bi-Maxwellian VDF, for the same value of the anisotropy. 

For lower values of the anisotropy, when the drift velocity becomes larger than the thermal speed, $U/\alpha_e>1$, as is the case of the middle right panels, $A=A_{\rm{eff}}=0.32$, we observe that for the counter-beaming electrons (lower panel) a secondary instability appears due to the Landau interaction provided by the separation of the two beams in velocity space (see Figure~\ref{vdf}(c)). This is the electrostatic two-stream instability (ETSI), also aperiodic (because counter-beams are symmetric), and reaches its maximum growth rate at lower angles and high wavenumbers, compared with the CBFI, but the later is still dominant with maximum growth rate $\gamma_{\rm{max}}/\Omega_e\approx0.77$ at $\theta\approx81^\circ$. Finally, for even lower anisotropy, $A=A_{\rm{eff}}=0.20$, the maximum growth rate in the case of counter-beaming VDF is given by the ETSI, with  $\gamma_{\rm{max}}/\Omega_e\approx 3.46$, at very low angles, $\theta\approx 2^\circ$. We observe that for $A << 1$ (i.e., large deviations from isotropy), both TAFI and CBFI reach high levels of growth rate for perpendicular propagation, i.e., $\theta=90^\circ$, and correspond to the aperiodic instabilities of the ordinary mode, well-known in the literature as Weibel instability triggered by temperature anisotropy~\citep{Weibel-1959}, and filamentation instability driven by the counter-beaming populations~\citep{Fried-1959}.

The growth rates  displayed and color codded in Figure~\ref{fixed-aniso} are obtained 
for the same anisotropy $A_{\rm{e}} = A_{\rm{\rm{eff}}} = 0.41$, corresponding 
to $U/\alpha_e = 0.85$, and different beta values, for both instabilities, 
TAFI (upper panels) and CBFI (lower panels). For both cases the value of electron 
beta parameter increases from left to right. The first panels (left) show the case 
for $\beta_e \simeq 1.79$, near the threshold of TAFI, with very low growth rates
$\gamma/\Omega_e \sim 10^{-4}$, while those of CBFI may reach maximum values 
three orders of magnitude higher. This difference is progressively reduced with increasing the electron beta. For instance, in the second case, for $\beta_e \simeq 3.20$, TAFI reaches a maximum growth rate $\gamma_{\rm{max}}/\Omega_e \simeq 0.07$ at $\theta \simeq 65^\circ$ and $ck/\omega_{pe} \simeq 0.66$, while in the bottom panel for CBFI we obtain $\gamma_{\rm{max}}/\Omega_e \simeq 0.30$ at $\theta \simeq 73^\circ$ and $ck/\omega_{pe} \simeq 0.69$. In the last case, for hotter (or less magnetized) electrons with $\beta_e =10.0$, the peaking growth rate of TAFI $\gamma_{\rm{max}}/\Omega_e \simeq 0.33$  is found at $\theta \simeq 79^\circ$ and $ck/\omega_{pe} \simeq 0.70$, and that for CBFI $\gamma_{\rm{max}}/\Omega_e \simeq 0.60$ at $\theta \simeq 80^\circ$ and $ck/\omega_{pe} \simeq 0.58$. In this case the thermal spread of counter-beaming electrons is sufficiently large to reduce not only the effective anisotropy, but also the contrast with the temperature anisotropy, corresponding to the first cases in Figure~\ref{vdf}. 

\begin{figure*}[ht!] 
  \begin{center}
    \includegraphics[width=0.95\textwidth]{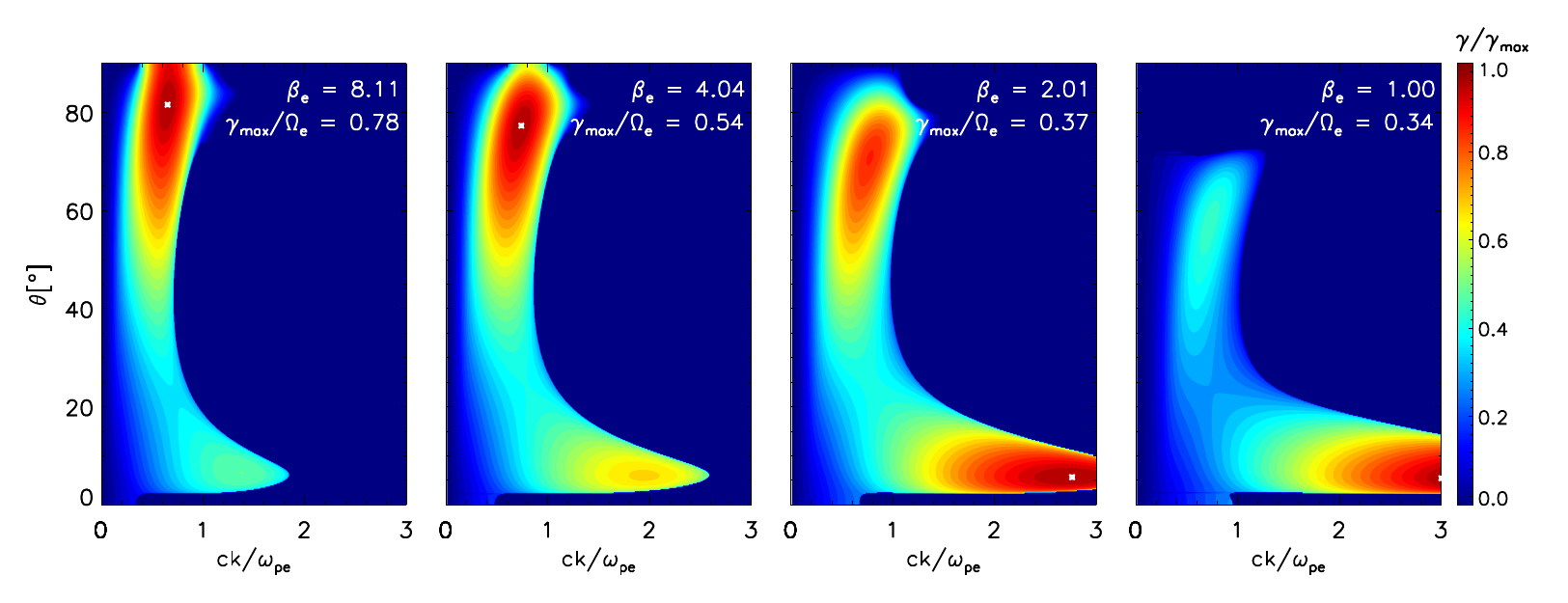}
    \caption{\label{BEFHI2ETSI} Growth rate as function of wave number and propagation 
		angle for symmetric counter-beams with $U=\alpha_e$ (or $A_{\rm{\rm{eff}}} = 1/3$), 
		and different, decreasing values of plasma beta. For better display, in each panel 
		growth rates are normalized to the maximum growth rate.}
  \end{center}
\end{figure*}

\section{Interplay and competition of CBFI with ETSI} 
\label{etsi}

As seen in the previous sections, numerical results shown that our instabilities, TAFI and CBFI, manifest not only similarities but even more contrasting properties. Thus, \cite{Lopez2020} reported the existence of those limit conditions of counter-beaming electron plasmas, susceptible to CBFI and another instability of different nature, namely, the electrostatic two-stream instability (ETSI). These regimes of interplay (and competition) of CBFI with ETSI are of particular interest in applications, being specific to counter-beaming electrons only, and thus allowing them to be distinguished from anisotropic distributions with similar allures but due to temperature anisotropy $T_\parallel > T_\perp$. 

\begin{figure*}[ht!] 
  \centering
    \includegraphics[width=0.95\textwidth]{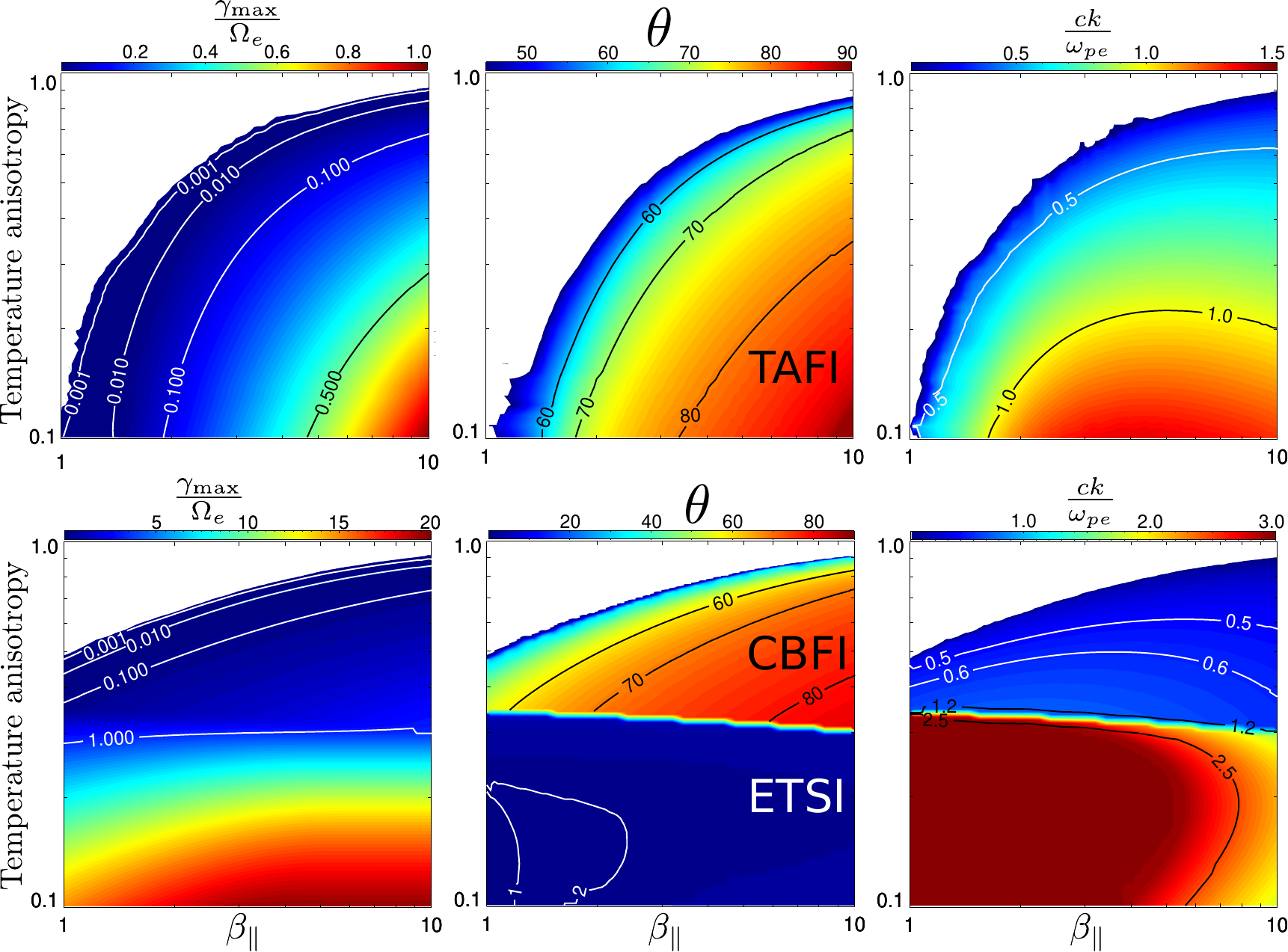}
    \caption{Diagrams of anisotropy vs. parallel beta for TAFI (upper panels) and CBFI (lower panels) showing contours of the maximum growth rates (thresholds, left panels), and contours of the propagation angle (middle panels) and wave-number (right panels) at maximum growth rate.}
    \label{fig:aniso-beta}  
\end{figure*}

The competition between CBFI and ETSI has already been discussed for the last two cases displayed in Figure~\ref{fixed-beta} (for a fixed $\beta_e = 7.06$), when the beaming speed is comparable or exceeds the thermal speed.
Exemplified in Figure~\ref{BEFHI2ETSI} are the unstable solutions obtained for other four distinct cases, corresponding to different, decreasing values (from left to right) of electron plasma beta.
Growth rates (normalized and color coded as indicated in the figure) are derived for symmetric counter-beams with beaming speed equal to their thermal speed, i.e., $U=\alpha_e$ (or $A_{\rm{\rm{eff}}} = 1/3$).
In the first panel (left), for large $\beta \simeq 8.11$ the peaking growth rate (indicated in the figure) is given by the CBFI at large propagation angle ($\theta\sim 82^\circ$) and small wave number ($ck/\omega_{pe} \simeq 0.65$). For lower beta the CBFI growth rates decrease and restrain to lower propagation angles, but the ETSI growth rates increase at quasi-parallel angles of propagation. Indeed, second panel of Figure~\ref{BEFHI2ETSI} shows that for $\beta=4.04$ the maximum growth corresponds to the CBFI at $\theta \sim 77^\circ$ and   
$ck/\omega_{pe} \simeq 0.7$, but also there is another (smaller) peak at $\theta\sim 6^\circ$ and $ck/\omega_{pe} \simeq 1.95$, corresponding to the ETSI. For the last two cases, when, e.g., $\beta_e =$ 2.01 and 1.00, the highest peaking (maximum) growth rates are associated with ETSI (at $\theta\sim 5^\circ$, and $ck/\omega_{pe} \simeq 2.78$, and $ck/\omega_{pe} \simeq 3.05$, respectively) which are thus predicted to develop faster than our CBFI.

\begin{table}
    \caption{Fitting parameters for Eq.~(\ref{eq:thres}).\label{t2}}
    \begin{tabular}{@{}lccc}
    \hline
    $\gamma_{\rm{max}}/\Omega_e$ & $\qquad a\qquad$ & $\qquad b \qquad$ &  $\qquad \beta_0\ \qquad$ \\
 \hline
 \multicolumn{4}{c}{ Temperature anisotropy firehose instability} \\
 \hline
          $10^{-3}$     & 1.478 & 0.668 & 0.193    \\   
        $10^{-2}$   & 1.372 & 0.940 & 0.206    \\
        $10^{-1}$    & 2.002 & 1.251 & 0.865    \\ 
    \hline
\multicolumn{4}{c}{ Counter beaming firehose instability} \\
 \hline
          $10^{-3}$     & 0.870 & 0.502 & 0.791    \\   
        $10^{-2}$   & 1.034 &0.836 & 1.068    \\
        $10^{-1}$    & 2.658 & 1.339 & 2.464   \\ 

    \hline 
    \end{tabular}\\
\end{table}

\section{Thresholds: TAFI vs. CBFI} \label{sec:thres}

\begin{figure}[t!] 
  \centering
    \includegraphics[width=0.9\textwidth]{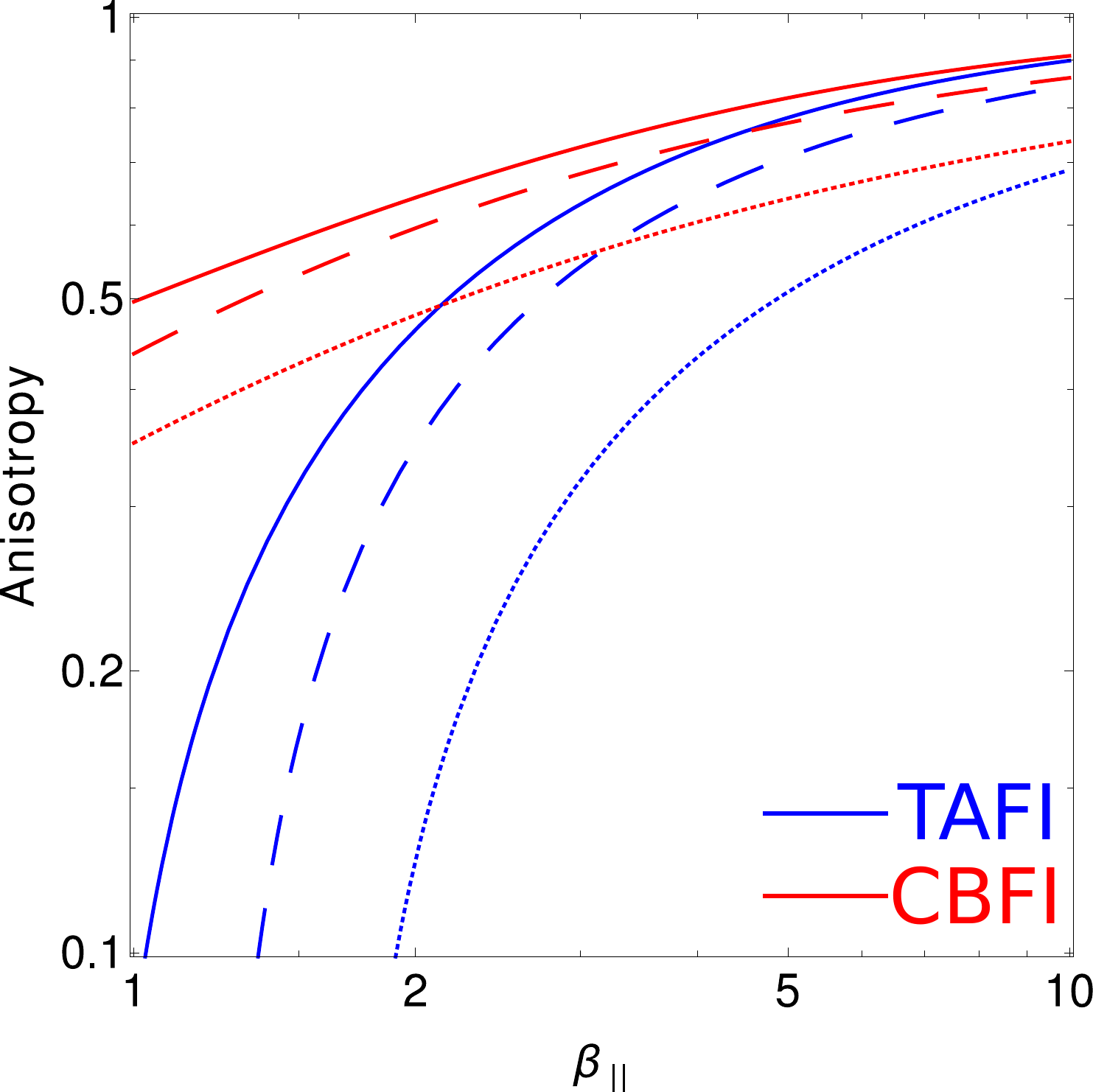}
    \caption{Instability thresholds in a diagram of anisotropy vs. parallel plasma beta derived as iso-contours of maximum growth rates (in units of the electron gyrofrequency): $\gamma_{\rm{max}}/\Omega_e$ = 10$^{-3}$ (solid), 10$^{-2}$ (dashed), and 10$^{-1}$ (dotted). Blue and red lines correspond to TAFI and CBFI instabilities, respectively. All lines are plotted using Eq.(\ref{eq:thres}), considering the values for the fitting parameters in Table \ref{t2}.}
    \label{fig:thresholds} 
\end{figure}

In this section we analyze the threshold conditions, presented by comparison in Figures~\ref{fig:aniso-beta} and \ref{fig:thresholds}, for both TAFI (upper panels) and CBFI (lower panels).
These thresholds provide a broad picture of the plasma system conditions favorable to our instabilities, in terms of the main plasma parameters, e.g., the electron anisotropy and  plasma beta, as well as details about the properties of unstable waves, e.g., the level of the maximum growth rate, the propagation angle, or the range of unstable wave numbers. 
The theoretical thresholds are confronted with the limits of quasi-stationary states reported by observations in space plasmas to elucidate the role played by instabilities in constraining kinetic anisotropies \citep{Stverak-etal-2008, Lazar-etal-2017, Shaaban-etal-2019b}. 

The maximum growth rates in the left panels in Figure~\ref{fig:aniso-beta} show comparable levels, and CBFI is predicted even for lower values of the parallel beta ($< 1$), where TAFI may not develop. However, for beams with higher speeds, or lower values of the effective anisotropy, e.g., $A_{\rm eff} < 0.4$, contour levels belong to ETSI with maximum growth rates superior to those of CBFI. This is more clearly indicated by either the contours of the propagation angles (middle panels), which make the difference between CBFI at highly oblique angles and the quasi-parallel ETSI, or by the contours of the (normalized) wave numbers (right panels), which differentiate between CBFI at low wave numbers and ETSI at larger wave numbers.

Contours of the instability thresholds given by the maximum growth rates as a function of anisotropy and parallel beta are quantified by fitting to the following law
\begin{equation}
\label{eq:thres}
    A = 1 - \frac{a}{(\beta_\parallel + \beta_0)^b},
\end{equation}
similar to other laws used in the analysis of firehose instabilities \citep{Gary-Nishimura-2003, Shaaban-etal-2018b}.
Given in Eq.~\ref{eq:thres} are the values obtained for the fitting parameters $a$, $b$ and $\beta_0$ for three levels of maximum growth rates $\gamma_{\rm{max}}/\Omega_e$ = 10$^{-3}$, 10$^{-2}$, and 10$^{-1}$. These are used in Figure~\ref{fig:thresholds} to better illustrate and compare the threshold conditions of our instabilities, TAFI (blue lines) and CBFI (red lines). From the Figure we can see that for larger beta values ($\beta_\parallel > 5$) and anisotropy $>$ 0.5 both instabilities exhibit similar thresholds contours, specially for the $\gamma_{\rm{max}}/\Omega_e$ = 10$^{-3}$, and 10$^{-2}$ levels. However, for more anisotropic cases (or smaller beta values) the instability thresholds for the TAFI and CBFI are quite different, being the CBFI more unstable for all plasma conditions. 

\section{Conclusions}
\label{sec:conclusions}

It has recently been shown, in theory and numerical simulations, that the aperiodic electron firehose mode can be destabilized not only by the temperature anisotropy ($T_\parallel > T_\perp$) of electrons, but also by the field-aligned counter-beaming electrons \citep{Lopez2020}. The electron firehose instabilities are intensively investigated in space plasmas, where particle-particle collisions are rare and plasma dynamics is expected to be controlled by the wave-particle interactions  \citep{Li-Habbal-2000, Gary-Nishimura-2003, Camporeale-Burgess-2008, Shaaban-etal-2019, Lopez-etal-2019, Saeed-etal-2017,Shaaban-etal-2018a, Shaaban-etal-2018b, Lopez2020}.
In the present paper we have compared these two instabilities, the temperature anisotropy firehose instability (TAFI) and the counter-beaming firehose instability (CBFI), by providing a comprehensive characterization of their similarities, but also their contrasting properties. This comparison was made for analogous plasma conditions, when the temperature anisotropy is comparable to the effective anisotropy of the counter-beaming electrons defined in Eq.~\ref{aeff}. 

Both instabilities are predicted for roughly the same interval of unstable wave numbers, and same range of highly oblique angles of propagation. Instead, the (maximum) growth rates of CBFI can reach higher values, up to one order of magnitude higher than those of TAFI. However, this diference is reduced with increasing plasma beta (e.g., if the anisotropy is constant), that means for a larger thermal spread of electron distributions. On the other hand, for the same plasma beta parameter but higher anisotropies, the CBFI can be competed by the ETSI, which is also aperiodic, but develops at quasi-parallel angles and larger wave numbers. This electrostatic instability has markedly different properties, which can be particularly useful in the observational analysis of velocity distributions. Thus, the presence of ETSI, e.g., in the regimes of interplay with CBFI, may help to identify the anisotropic distributions with counter-beaming electrons, and distinguish from those with similar profiles but given by a temperature anisotropy $T_\parallel > T_\perp$. However, for the situations when the electrostatic fluctuations are not detected, such a distinction can only be made by the contrasting properties of TAFI and CBFI. Synthesized in Figures \ref{fig:aniso-beta} and \ref{fig:thresholds}, these properties indicate a relevant particularity of CBFI, which, by contrast with TAFI, can be excited for the low plasma beta (i.e., $\beta < 1$) regimes. This is consistent with the results provided by~\citet{Hadi2014}, which showed that the presence of beams may introduce remarkable changes to the instability thresholds and make the plasma unstable for $\beta < 1$ and $T_\perp /T_\parallel < 1$ regime, a behaviour that is not specific for TAFI.

Under the context of space plasmas, which are collision-poor or even collision-less, our results may help to explain the regulation of the kinetic anisotropies of, e.g., solar wind plasma populations, and provide insights leading to realistic kinetic models of solar wind electrons. The instability thresholds may be useful to understand the differences between electron observations and theoretical predictions for, e.g., temperature anisotropy driven instabilities, as can be found in \citet{Stverak-etal-2008} or \citet{Adrian2016}, among others. Thus, only thresholds of TAFI are commonly applied to explain the observed limits of electron temperature anisotropy ($T_\perp /T_\parallel < 1$), although in high-speed winds instabilities triggered by the electron beam or strahl may also contribute with additional constraints. In particular, solar wind electrons observations usually exhibit good agreements with kinetic instability thresholds for $T_\perp /T_\parallel > 1$. For $T_\perp /T_\parallel < 1$, however, the thresholds for TAFI bound the data but not as close to the observed limits \citep[see e.g.][]{Stverak-etal-2008,Adrian2016}. Thus, the CBFI or ETSI may be relevant in order to obtain better agreement between theoretical predictions and observations. Future studies should therefore extend and refine the investigations of these instabilities, in particular, for space plasma conditions, and provide plausible explanations for the observations.

\section*{acknowledgements}
The authors acknowledge support from the Katholieke Universiteit Leuven, Ruhr University Bochum, and Mansoura University. These results were also obtained in the framework of the projects C14/19/089 (C1 project Internal Funds KU Leuven), G.0D07.19N (FWO-Vlaanderen), SIDC Data Exploitation (ESA Prodex-12), Belspo project B2/191/P1/SWiM, and Fondecyt No. 1191351 (ANID, Chile). P.S. Moya is grateful for the support of KU Leuven BOF Network Fellowship NF/19/001. R.A.L. acknowledges the support of ANID Chile through FONDECyT grant No. 11201048.


\bibliography{MS_new}{}
\bibliographystyle{aasjournal}



\end{document}